\documentclass[reprint,aip,rsi,amsmath,amssymb,nofootinbib]{revtex4-1}

\usepackage{graphicx}
\usepackage{dcolumn}
\usepackage{bm}

\usepackage{float}
\usepackage{caption}
\usepackage{subcaption}
\usepackage{xcolor}
\usepackage{soul}
\usepackage{appendix}
\usepackage{chngcntr}
\usepackage{multirow}
\usepackage{textcomp, gensymb}

\begin{document}
\title{Evidence of Gas Phase Nucleation of Nano Diamond in Microwave Plasma Assisted Chemical Vapor Deposition}

\author{Tanvi Nikhar}
\email{nikharta@msu.edu}
\affiliation{Department of Electrical and Computer Engineering, Michigan State University, 428 S. Shaw Ln., East
Lansing, MI 48824, USA}

\author{Sergey V. Baryshev}
\email{serbar@msu.edu}
\affiliation{Department of Electrical and Computer Engineering, Michigan State University, 428 S. Shaw Ln., East
Lansing, MI 48824, USA}
\affiliation{Department of Chemical Engineering and Materials Science, Michigan State University, 428 S. Shaw Ln., East
Lansing, MI 48824, USA}

\begin{abstract}
The mechanism of ballas like nano crystalline diamond formation (NCD) still remains elusive, and this work attempts to analyze its formation in the framework of activation energy ($E_\text{a}$) of NCD films grown from H$_\text{2}$/CH$_\text{4}$ plasma in a 2.45 GHz chemical vapor deposition system. The $E_\text{a}$ is calculated using the Arrhenius equation corresponding to the thickness growth rate
while using substrate temperature ($\sim1000-1300$ K) in all the calculations. While the calculated values match with the $E_\text{a}$ for nano diamond formation throughout the literature, these values of $\sim$10 kcal/mol are lower compared to $\sim$15 -- 25 kcal/mol for standard single crystal diamond (SCD) formation, concluding thus far, that the energetics and processes involved are different. In this work, To further investigate this, the substrate preparation and sample collection method are modified while keeping the growth parameters constant. Unseeded Si substrates are physically separated from the plasma discharge by a molybdenum plate with a pinhole drilled in it. Small quantities of a sample substance are collected on the substrates. The sample is characterized by electron microscopy and Raman spectroscopy confirming it to be nano diamond, thus, suggesting that nano diamond self nucleates in the plasma and flows to the substrate which acts as a mere collection plate. It is hypothesized then, if nano diamond nucleates in gas phase, 
gas temperature has to be used in the Arrhenius analysis. The $E_\text{a}$ values for all the nano diamond films are re-calculated using the simulated gas temperature ($\sim1500-2000$ K) obtained from a simple H$_\text{2}$/CH$_\text{4}$ plasma model, giving values within the range characteristic to SCD formation. A unified growth mechanism for NCD and SCD is proposed concluding that the limiting reactions for NCD and SCD formation are the same.

\end{abstract}

\maketitle

\section{\label{sec:Introduction}Introduction}

The growth of diamond through chemical vapor deposition (CVD) is reported extensively on an existing diamond structure, such as on a diamond substrate or on a diamond seed (achieved by various seeding processes performed on a non-diamond substrate), and a number of studies have been done hypothesizing the mechanism of adding C atoms from source gas to the base diamond structure that results in growth \cite{Fedoseev1975, Harris1990_MethylRadical, Harris1990, Frenklach1991_MolecularProcesses, Butler1993, Wolden1994_ReducedRn}. The precipitation of solid diamond from gaseous carbonaceous species without having an already defined diamond structure to grow on has also been reported \cite{Spitsyn1981_c-Substrates, Maeda1992_Ceramic, Pehrsson1992_GraphiteFibers, Kobayashi1993_Fe-Si, Chai1994_cBN} but analysis of the mechanism behind it is far less analyzed. To initiate growth without an existing diamond structure to replicate on, the reactive species from gas, at some point, need to form the first diamond structure -- a nucleus, on which the replication (epitaxy) process may follow to form a diamond crystal. 

Without an existing diamond seed, this nucleation process is mainly understood to be originating at favorable nucleation sites on the surface of a substrate, such as on defects introduced by scratching the substrate with abrasives (including non-diamond materials that do not leave a diamond seed behind), carbide forming substrates, etc. \cite{Ch6_Book_Nucleation} For mirror-polished surfaces, a commonly used technique is bias-enhanced nucleation, where charged ions impinge upon the surface implanting C atoms in the upper layers of the substrate and these act as active sites for nucleation \cite{Yugo1991, BEN_Garcia2000, Science_Lifshitz2002}. Hence, diamond nucleation is widely understood to be a surface phenomenon.


In low-pressure CVD systems though, gas phase nucleation of diamond has been rarely discussed, but there are some reports, however scarce, discussing its possibility based on experimental observations. Mitura $et$ $al.$ \cite{Mitura1987} observed agglomerates of diamond particles separate from the rest of the cohesive film deposited in an RF methane plasma and concluded that these particles must have formed in the gas phase, separate from the film formed on the substrate surface. Howard $et$ $al.$ \cite{Howard1990} reported the synthesis of diamond powder in the gas phase by microwave-assisted combustion of acetylene in oxygen at low pressures. Frenklach $et$ $al.$ \cite{Frenklach1989} were successfully able to obtain diamond downstream in a low-pressure CVD reactor with mixtures of dichloromethane and oxygen hence definitely confirming that nucleation does not necessarily require a surface and is happening in the gas phase, although tedious post-treatment of the collected sample was required to reveal the presence of diamond grains. They were also able to achieve induced nucleation of diamond by adding diborane to a mixture of acetylene, hydrogen and argon.\cite{Frenklach1991}

Various theoretical studies as reviewed by Liu $et$ $al.$ \cite{Ch4_Book_Nucleation}, have reported on the plausibility of gas phase nucleation of diamond. However, the high nucleation density necessary for the formation of a continuous film can only be attributed to surface nucleations. Hence, the importance of gas phase nucleation is not well understood and is recognized at best as an accompanying process to processes taking place on the substrate surface. Hwang $et$ $al.$ \cite{Hwang1996} presented a theoretical confirmation of gas phase nucleation based on the study of carbon atom potential. But there is not enough experimental proof seen in a standard low-pressure hydrogen methane plasma CVD. 

The hypotheses on the diamond formation mechanism via CVD, include a number of proposed reactions 
and models of the growth process to investigate the energies involved at various steps of these reactions.\cite{Wang1991_CyclicDeposition, Harris1991_FilamentAsst,  Frenklach1992_MonteCarlo,  Garrison1992, Frenklach1994_ChemicalRnMechamisms, Kang2000, Cheesman2008} A common thread in all these studies is the treatment of the growth process as a surface phenomenon. 
Gas kinetics have mostly been studied separately from surface kinetics because combining the two would immensely increase the complexity of the system to be modeled.\cite{SeparateModels_Frenklach1991} These calculations are done based on the assumption that surface kinetics leading to deposition can be considered independent of gas kinetics, which are then used to thermodynamically warrant the proposed mechanisms. This leaves out the possibility of important processes occurring in the gas phase that may end up altering the overall kinetics of the process. It begs the question of whether the growth kinetics can be treated as an independent  function of surface phenomena.

The commonly accepted mechanism for the growth of diamond ($sp^3$ phase of carbon) is by CH$_\text{3}$ radical incorporation. The activation energy calculated through experimentally obtained data for diamond films with varying grain sizes reveals that for larger grains having faceted morphology as seen in microcrystalline diamond films (MCD), the activation energy is $\sim$10 kcal/mol. \cite{Brazil2014} This energy matches up with the energy required for CH$_\text{3}$ incorporation into the lattice, thus, making it the rate-limiting step for such type of diamond growth. On the other hand, for smaller crystallites of diamond having ballaslike morphology as seen in NCD films, the activation energy was found to be consistently 
lower, $\sim$6 kcal/mol. \cite{Brazil2014} Thus, the formation of ballaslike diamond morphology was hypothesized to be different from the canonical CH$_\text{3}$ incorporation mechanism. The reason for the lowering of the activation energy was not clear and was attributed to various possibilities: competition of diamond grain with non-diamond phase growth (especially graphitic); a mechanism led by C$_\text{2}$ dimer incorporation into the lattice which corresponds to $\sim$6 kcal/mol; high renucleation rate that probably interrupts $sp^3$ grain growth by imposing another rate-limiting step that is also unknown.

What is concluded so far is that faceted and ballaslike morphologies must follow different reaction mechanisms owing to their different activation energies of formation. In this work, we investigate further on this difference to get a better understanding of the mechanism for the formation of ballaslike morphology. The activation energies of ballaslike NCD films produced in our previous work \cite{Dynamic_Graphitization} are calculated from the Arrhenius equation used to approximate the relation of growth rate 
as a function of temperature. New syntheses is carried out by depositing ballaslike NCD using two distinct methods; in one of which plasma was decoupled from the substrate. From this, a hypothesis of gas phase nucleation of diamond from hydrogen methane plasma in a conventional low-pressure CVD system is formulated. Aided by a basic computational model describing the H$_\text{2}$/CH$_\text{4}$ microwave assisted CVD plasma, a correction in the calculation of activation energy for the grown films is proposed. The recalculated energy values are then used to discuss inferences about the growth mechanism involved in ballaslike morphology formation.


\section{\label{sec:Samples}Samples}

The E$_\text{a}$ calculations in this study are done for the NCD films deposited on intrinsic Si(100) substrates in a 2.45 GHz microwave plasma assisted chemical vapor deposition (MPACVD) system.\cite{DiamondFilmsHandbook} The films were comprehensively described and characterized in our previous work. \cite{Dynamic_Graphitization} We particularly focus on the films synthesized in pure H$_\text{2}$/CH$_\text{4}$ plasma denoted by the SA series corresponding to 0\% nitrogen summarized in Table I of Ref. \onlinecite{Dynamic_Graphitization}.



All samples were characterized with a 532 nm probing laser on a Horiba Raman spectrometer. High-resolution scanning electron microscopy (SEM) was performed using a JEOL JSM 7500F to study the surface morphology of the samples. The Si substrates are weighed before and after deposition using a Mettler Toledo XS 105 balance with 10$^\text{-5}$ g readability to find the weight of the deposited material given by $\Delta w$. 

\section{\label{sec:Results}Activation Energy Results}

\subsection{Thickness calculation}
The Raman spectra of the NCD films presented in our previous work\cite{Dynamic_Graphitization} reveal that as the substrate temperature, $T_s$ increases, the D peak shifts from 1333 cm$^\text{-1}$ to 1350 cm$^\text{-1}$ and the G peak gets narrower and shifts towards 1590 cm$^\text{-1}$. The evident phase transformation of the material from ultra-nano-crystalline diamond to nano-crystalline graphite is used here to approximate the thickness of the deposited material using the following relation:
\begin{equation}
t  = \frac{\Delta w \cdot 10^4}{S_a\cdot \rho}~\text{in}~[\mu\text{m}]
\label{eq:thickness}
\end{equation} 
where, $\Delta w$ = weight difference before and after deposition (g), $S_a$ = surface area of the film (cm$^\text{2}$), and\\
$\rho$ = 
$\begin{cases} 
\rho_{dia}, & \text{if the D peak is at 1333 cm$^\text{-1}$} \\ 
\rho_{gra}, & \text{if the D peak is shifted to 1350 cm$^\text{-1}$} 
\end{cases}$
where, $\rho_{dia}$ = 3.51 g/cm$^3$ is the density of pure diamond, and $\rho_{gra}$ = 2.27 g/cm$^3$ is the density of pure graphite.

The growth of the films is interpreted in terms of 
thickness $t$
. Since all the films are deposited over a period of 1 hour, the growth rate is given by $k$  = thickness/time = $t [\mu\text{m}]$/hour.

\subsection{Activation energy calculation}

The activation energy 
for the growth 
is calculated using the Arrhenius equation, given as:
\begin{equation}
k = A \cdot e^{-E_{a}/RT}
\label{eq:arrhenius}
\end{equation}
Hence,
\begin{equation}
E_{a} = -RT \hspace{5pt} ln(k/A)
\label{eq:activation_energy}
\end{equation}
where, $R = 8.31$ J/mol$\cdot$K = $8.31\times 0.000239$ kcal/mol$\cdot$K is the ideal gas constant, $T$ = 
temperature (in Kelvin), $k$ = rate constant
, and $A$ = pre-exponential factor calculated from the Arrhenius plot: ln($k$) vs 1/$T$ whose intercept is ln($A$).\\


The term $k$ in the expression of $E_a$ is the rate constant of the reaction taking place at temperature $T$. Here, this temperature corresponds to the reactions that lead to the deposition of the film. A common assumption throughout literature so far has been that since the deposition of the nanodiamond particles happens on the substrate surface which eventually builds up to form the NCD film, the reactions involved in the formation of these particles must happen on the surface itself. Thus, in all the studies, the temperature of the substrate measured experimentally is used to calculate the activation energy for the formation of the films. 

\begin{figure}[]
        \includegraphics[width =0.45\textwidth]{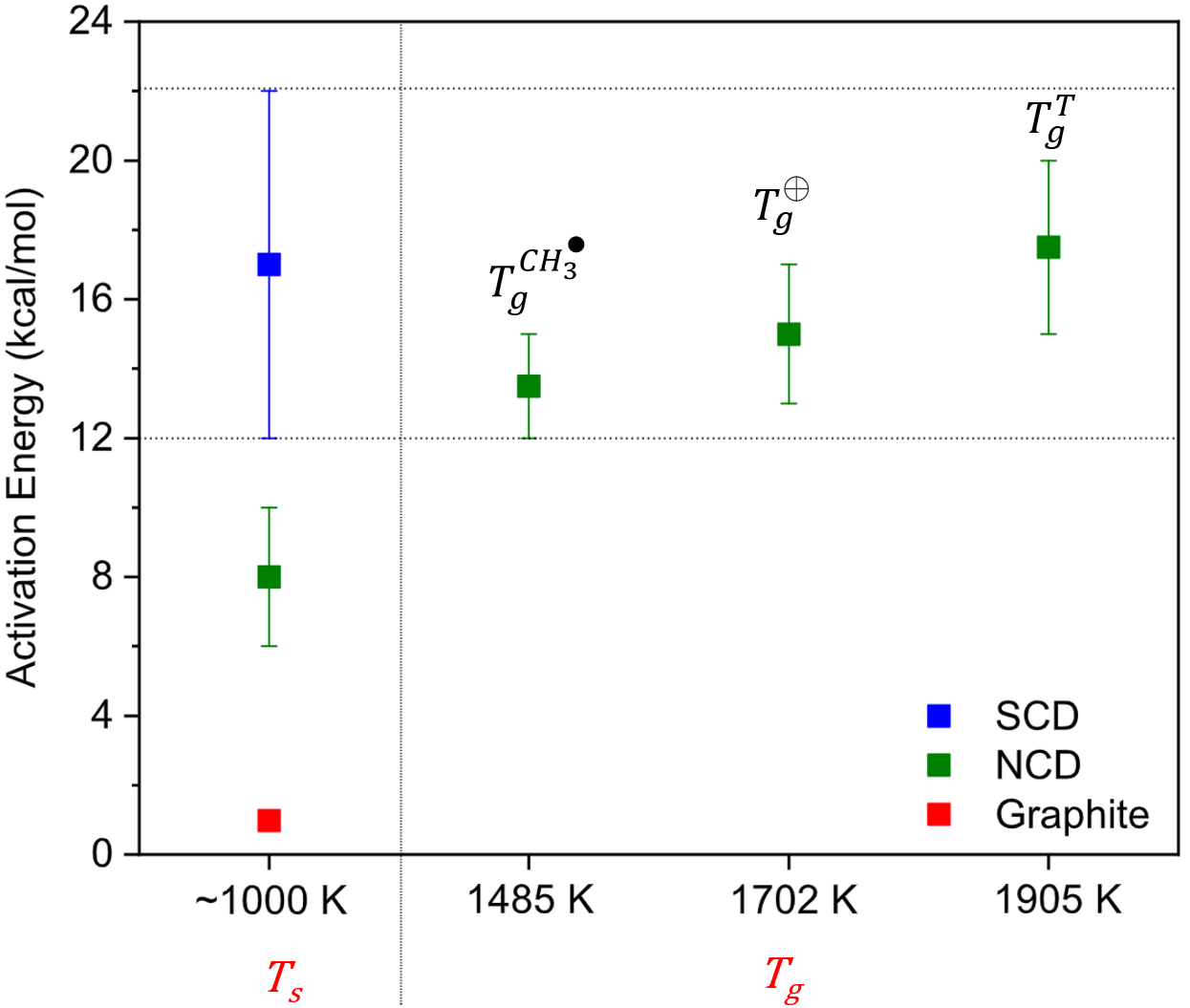}
\caption{Activation energies 
of NCD films calculated using substrate temperature ($T_s$) and gas temperature ($T_g$) as compared with the activation energy of formation of SCD and pure graphite.} 
\label{fig:Ea}
\end{figure}

For our experiments, the temperature of the substrate $T_s$ was measured using an infrared pyrometer, the values of which are reported as deposition temperature in Table I in Ref. \onlinecite{Dynamic_Graphitization}. The activation energies calculated using Eq.~\ref{eq:activation_energy} with $T=T_s$ 
are summarized in Fig.~\ref{fig:Ea} above the label $T_s$ at the 1,000 K mark. Activation energies for the formation of pure graphite\cite{Fedoseev1979} and SCD\cite{Kondoh1993,Kang2000} are also plotted for comparison. This result agrees extremely well with the results of the previous meta-study\cite{Brazil2014} that summarized and analyzed data on nanodiamond obtained by different deposition techniques. 
Since $E_a$ for nanodiamond corresponding to $T_s$ has consistently been found lesser than that for SCD, the mechanisms for the formation of NCD and SCD have been assumed to be different. As summarized by Barbosa $et$ $al.$, \cite{Brazil2014} the rate-limiting reaction for SCD formation is understood to be CH$_\text{3}^{\bullet}$ incorporation while the limiting reaction for NCD is not known yet.

The formation of nanodiamond films is attributed to a high renucleation rate. Based on their observation of an orange glow on the edges of the plasma discharge, May $et$ $al.$\cite{May2010} suggested that the high renucleation rate could be due to particulates present in the outer edge of the plasma. These particulates are speculated to act as nucleation or defect sites on the growth surface promoting film growth and a high renucleation rate. A similar orange glow was also observed in the discharges produced in our previous work on nanodiamond synthesis.\cite{Dynamic_Graphitization} 

NCD synthesis on unseeded Si substrate was attempted to find whether self-seeding (from the afore-mentioned particulats) and further coalescence and continuous film growth was attainable. The results as presented in Ref. \onlinecite{Dynamic_Graphitization} showed that the first hour of deposition resulted in the formation of colonies mostly separate from each other. After the second hour, some of the colonies were found to merge together towards the formation of a continuous film. 
SEM images and Raman spectra of these colonies confirmed that the samples obtained on seeded and unseeded substrates were identical. The synthesis runs were repeated on seeded and unseeded Mo substrates, each producing identical samples to the ones on Si substrates. 

These results provided evidence of self-seeding, independent of the type of substrate or the substrate pre-treatment process. These results corroborated the previous hypotheses of diamond gas phase nucleation, where the seeding diamond particulates are formed in the plasma itself independent of the substrate.  If true, then the temperature term in Eqs.~\ref{eq:arrhenius}~and~\ref{eq:activation_energy} should be taken as the gas temperature $T_g$ and not the substrate temperature $T_s$.

\section{\label{sec:Experiment}Experiment}

\begin{figure}[h]
\centering
       \includegraphics[width=0.47\textwidth]{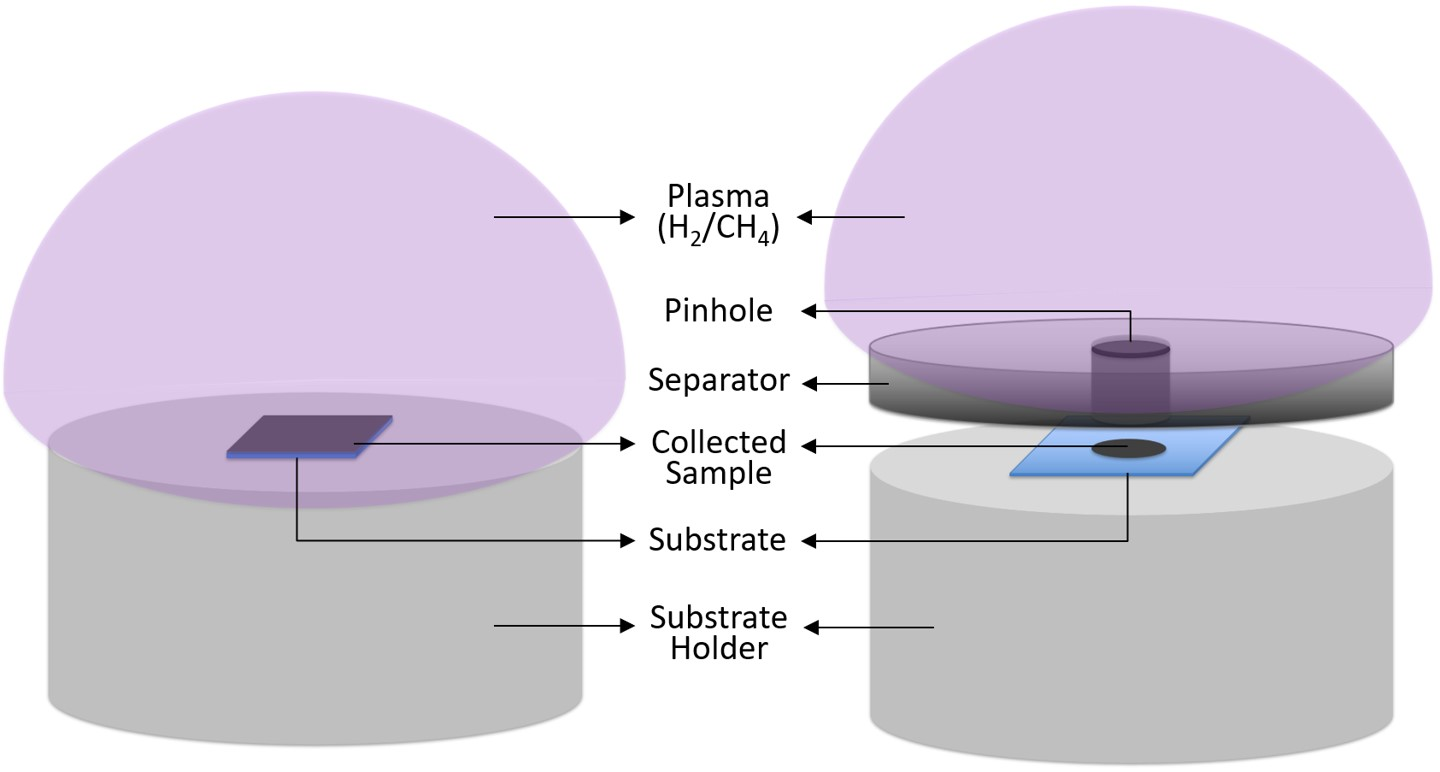}
\caption{Schematics showing the two setup arrangements used for the deposition of NCD samples: (left) Original setup: the samples are collected on a substrate directly exposed to the plasma; (right) Altered setup: the samples are collected through a pinhole in a Mo stub used to physically separate the plasma from the substrate.}
\label{fig:Setup}

\end{figure}

To confirm this hypothesis, in the present work, we further modified our original setup (Fig.~\ref{fig:Setup} (left)) to physically separate the plasma from the substrate by placing a Mo disc on top of the substrate removing any direct contact with the plasma as shown in Fig.~\ref{fig:Setup} (right). An unseeded Si substrate of $\sim$1~cm~$\times$~1~cm$~\times$~0.3~mm was placed in a pocket of the Mo substrate holder on top of which a Mo disc of diameter 7 cm and thickness 3.31 mm with a pinhole 
was placed with the intention of allowing any particles forming in the plasma to pass through the pinhole onto the substrate while keeping it physically separate from the plasma. 

The diameter of the disc is chosen wide enough so that the entire plasma is contained on top of it and does not flow off its edge, thus keeping the shape of the plasma unperturbed. 
The diameter of the pinhole is chosen in accordance with the debye length of the plasma near the surface of the disc,
ensuring that the plasma do not reach down to the substrate, hence maintaining physical separation.

{\begin{figure}[]
\centering
\includegraphics[width =0.45\textwidth]{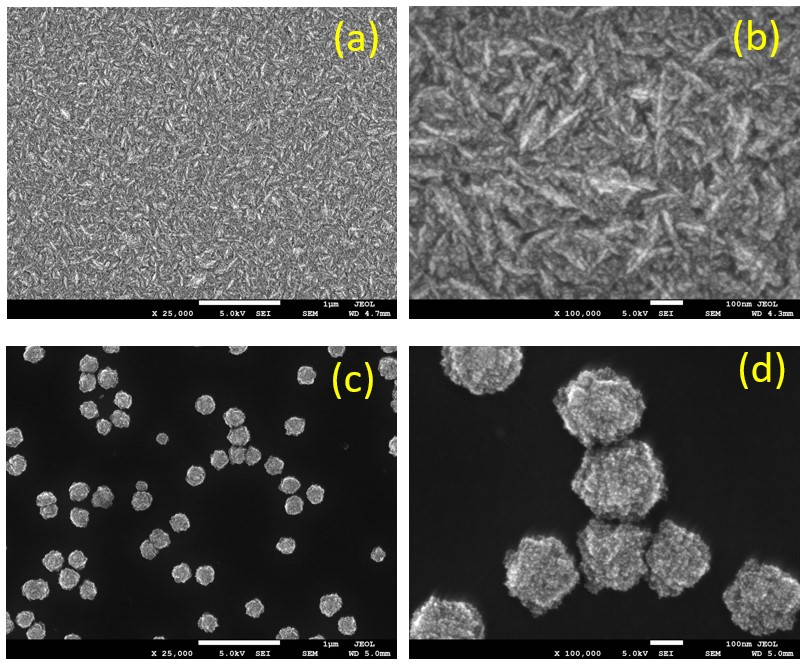}
\caption{SEM images of sample collected (a), (b) in the original setup as compared to samples collected in the altered setup through the pinhole (c), (d).} 
\label{fig:SEM_old_vs_new}

\centering
\includegraphics[width =0.5\textwidth]{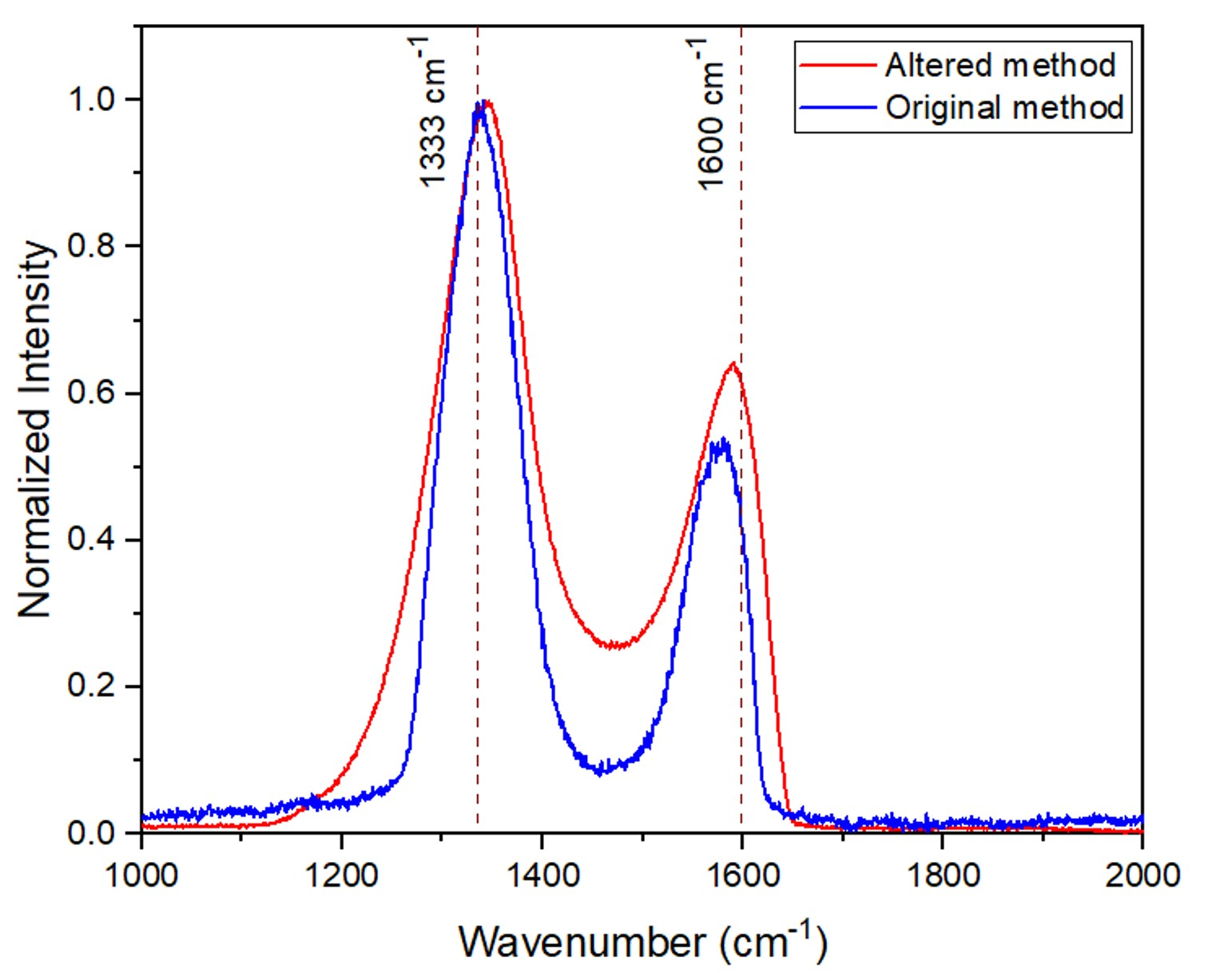}
\caption{Raman spectra comparing samples collected by the original and altered methods, both showing D and G bands typical to nanodiamond. 
} 
\label{fig:Raman_old_vs_new}
\end{figure}

Samples are collected with both setups using process conditions:
a hydrogen-rich H$_2$/CH$_4$ mixture was used with the CH$_4$ flow rate was kept at 10 sccm (5\% by volume), while the flow rate of H$_2$ was maintained at 200 sccm.

With the original setup leading to in-plasma collection, using seeded Si as the substrate, a continuous film of thickness $\sim$500 nm was formed over a deposition duration of 1 hour. The SEM images and Raman spectrum shown in Fig.~\ref{fig:SEM_old_vs_new}(a),(b) and \ref{fig:Raman_old_vs_new} respectively, present characteristics of the film typical for nanodiamond. With the new setup using unseeded Si as the substrate, the out-of-plasma sample collection through the pinhole was only observed after a deposition duration of 4 hours, exclusively over the substrate area exposed to the pinhole opening. A very low yield of the sample is obtained 
having poor adherence to the substrate and can be easily removed by a cotton swab. (The films obtained through conventional in-plasma deposition adhere strongly to the substrate.) The SEM analysis of the sample collected through the pinhole revealed the presence of individual colonies scattered only under the area of the pinhole while the rest of the substrate remains clear of any sample. The diameter of each colony ranges from 118 to 262 nm with the morphology as shown in Fig.~\ref{fig:SEM_old_vs_new}(c),(d) identical to the nanodiamond film shown in Fig.~\ref{fig:SEM_old_vs_new}(a),(b). The Raman spectrum shown in Fig.~\ref{fig:Raman_old_vs_new} in red is also similar to that of the original setup shown in blue.

The obtained results even further support the gas phase nucleation of nanodiamond formation. It indeed calls for replacing $T_s$ with $T_g$ when calculating the activation energy in Eq.~\ref{eq:activation_energy}. To obtain $T_g$ parameter, a model of the reactor was built to estimate the basic plasma parameters. 

\section{\label{sec:plasma_model}Numerical model of the plasma}

\subsubsection{Model formulation}

Numerical modeling of microwave CH$_\text{4}$/H$_\text{2}$ plasma was carried out in COMSOL with consideration of 2D computational domain with an axisymmetric problem statement. The corresponding symmetry axis crosses the center of the quartz dome and the substrate holder at their centers.

CH$_\text{4}$/H$_\text{2}$ plasma chemistry modeling is based on the solution of species and momentum conservation equations with drift-diffusion approximation for electrons and positive ions. The electron conservation equation is written as:
\begin{equation}
     \frac{\partial}{\partial t}(n_e) + \nabla\bm{\cdot}\vec{\Gamma_e} = R_e
    \label{eq:t}
\end{equation}
where $n_e$ is the electron density ($m^{-3}$). $\vec{\Gamma_e}$ is the electron flux vector derived from the momentum conservation equation and is given by: 
\begin{equation}
     \vec{\Gamma_e} = - n_e \mu_e \vec{E} - D_e \nabla n_e 
\end{equation}
where $\mu_e$ and $D_e$ are the electron mobility ($m^2/Vs$) and diffusion ($m^2/s$) constants respectively. $\vec{E}$ is the electric field. Suppose that there are $M$ reactions that contribute to the growth/decay of electron density, then the electron source term $R_e$ is given by:
\begin{equation}
    R_e = \sum_{j=1}^{M} x_j k_j N_n n_e
\end{equation}
where $x_j$ is the mole fraction of the target species for reaction $j$, $k_j$ is the rate constant ($m^3/s$) for reaction $j$, and $N_n$ is the total neutral density ($m^{-3}$).
\begin{equation}
    \rho_i \frac{\partial w_i}{\partial t} = \nabla \bm{\cdot} \vec{\Gamma}_i + R_i
\end{equation}

The electron energy balance equation solved to calculate electron temperature is given by:
\begin{equation}
    \frac{\partial}{\partial t}(n_\varepsilon) + \nabla \bm{\cdot} \vec{\Gamma}_\varepsilon + \vec{E} \bm{\cdot} \vec{\Gamma}_e = S_{en} - (\vec{u_e} \bm{\cdot} \nabla)n_e + Q/q
\end{equation}
where $n_\varepsilon$ is the energy density ($V/m^3$). $\vec{\Gamma}_\varepsilon$ is the electron energy flux given by:
\begin{equation}
\vec{\Gamma}_\varepsilon = -\mu_\varepsilon n_\varepsilon \vec{E} - D_\varepsilon \nabla n_\varepsilon
\end{equation}
where $\mu_\varepsilon$ is the electron energy mobility ($m^2/Vs$), $D_\varepsilon$ is the electron energy diffusivity ($m^2/s$). Suppose that there are $P$ inelastic electron-neutral collisions, then the energy loss/gain due to inelastic collions $S_en$ ($V/m^3s$) is can be given by:
\begin{equation}
    S_e = \sum_{j=1}^{P} x_j k_j N_n n_e \Delta \varepsilon_j
\end{equation}
where $\Delta \varepsilon_j$ is the energy loss from reaction j ($V$). $\vec{u_e}$ is the elecron drift velocity vector, and $Q$ is the heat source ($W/m^3$).

Ionization, electronic and vibrational state excitation of H$_\text{2}$, and dissociation cross-sections of H$_\text{2}$ and CH$_\text{4}$ \cite{Mankelevich2008, Hassouni2010, Yoon2008} were specified to get an accurate calculation of electron temperature. Since CH$_\text{4}$ presence in the mixture is 5\%, it was assumed that electron kinetics and microwave plasma properties are dominated by electron collisions with H and H$_\text{2}$.

\begin{table}[]

\renewcommand{\arraystretch}{1.5}
\begin{tabular}{c||c}
Volume Reaction         & Description                       \\
\hline
\hline
H$_\text{2}$ + e → H$_\text{2}$ + e       & Elastic collisions                \\
\hline
H + e  → H + e  &  Elastic collisions \\
\hline
H$_\text{2}$ + e → H$_\text{2}$(V) + e   & 1-3 vibrational levels excitation \\
\hline
H$_\text{2}$ + e → H$_\text{2}$ + e     & Electronic levels excitation      \\
\hline
H$_\text{2}$ + e → H$_\text{2}^\text{+}$ + 2e     & Direct ionization                 \\
\hline
H$_\text{2}$ + e → 2H + e        & Dissociation                      \\
\hline
H + e → H$^\text{+}$ + 2e         & Direct ionization                 \\
\hline
H$_\text{2}^\text{+}$ + H$_\text{2}$ → H$_\text{3}^\text{+}$ + H & H$_\text{3}^\text{+}$ ions formation              \\
\hline
H$_\text{3}^\text{+}$ + H → H$_\text{2}^\text{+}$ + H$_\text{2}$   &   \\
\hline
H$_\text{3}^\text{+}$ + e → 3H   &   \\
\hline
H$_\text{3}^\text{+}$ + e → H$_\text{2}$ + H   &   \\
\hline
H$_\text{2}^\text{+}$ + e → 2H   &   \\
\hline
2H$_\text{2}$ → 2H+H$_\text{2}$   &   \\
\hline
H$_\text{2}$+H → 3H   &   \\
\hline
2H+H$_\text{2}$ → 2H$_\text{2}$   &   \\
\hline
3H → H$_\text{2}$ + H   &   \\
\hline
CH$_\text{4}$ +H → CH$_\text{3}$ + H$_\text{2}$   &   \\
\hline
CH$_\text{3}$ + H$_\text{2}$ → CH$_\text{4}$ + H   &   \\
\hline
CH$_\text{3}$ + H → CH$_\text{2}$ + H$_\text{2}$   &   \\
\hline
CH$_\text{2}$ + H$_\text{2}$ → CH$_\text{3}$ + H   &   \\
\hline
CH$_\text{2}$ + H → CH + H$_\text{2}$   &   \\
\hline
CH+H$_\text{2}$ → CH$_\text{2}$ + H   &   \\
\hline
CH + H → C + H$_\text{2}$   &   \\
\hline
C + H$_\text{2}$ → CH$_\text{2}$ + H   &   \\
\hline
CH$_\text{3}$ + C → C$_\text{2}$H$_\text{2}$ + H   &   \\
\hline
C$_\text{2}$H$_\text{2}$ + H → CH$_\text{3}$ + H$_\text{2}$   &   \\

\end{tabular}
\caption{Summary of the gas phase reactions.}
\label{table:reactions}
\end{table}

\begin{figure*}[t]
\centering

        \begin{subfigure}{0.32\textwidth}
        \includegraphics[width=\linewidth]{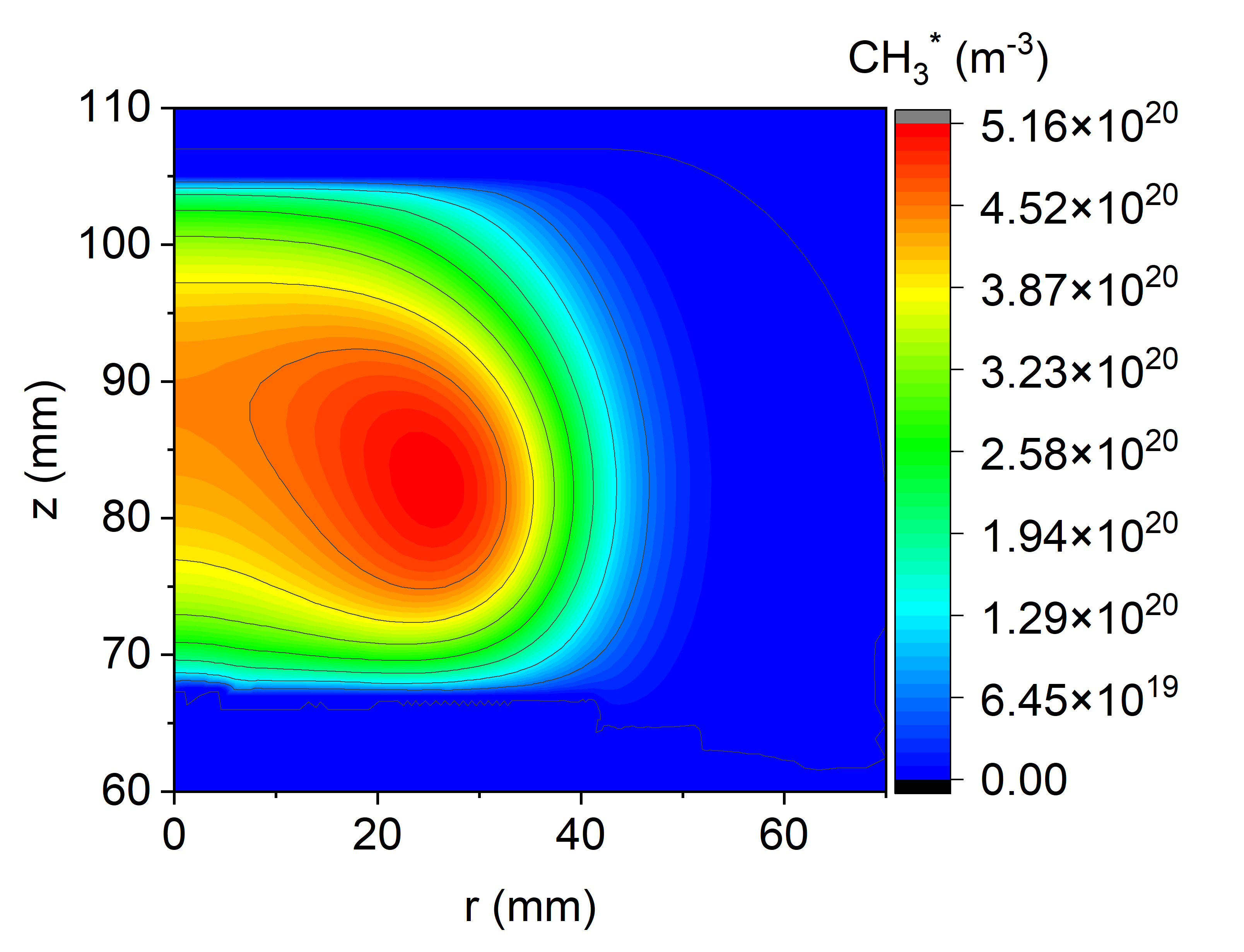}
         \caption{CH$_\text{3}$  density distribution}
    \end{subfigure}
    \begin{subfigure}{0.32\textwidth}
        \includegraphics[width=\linewidth]{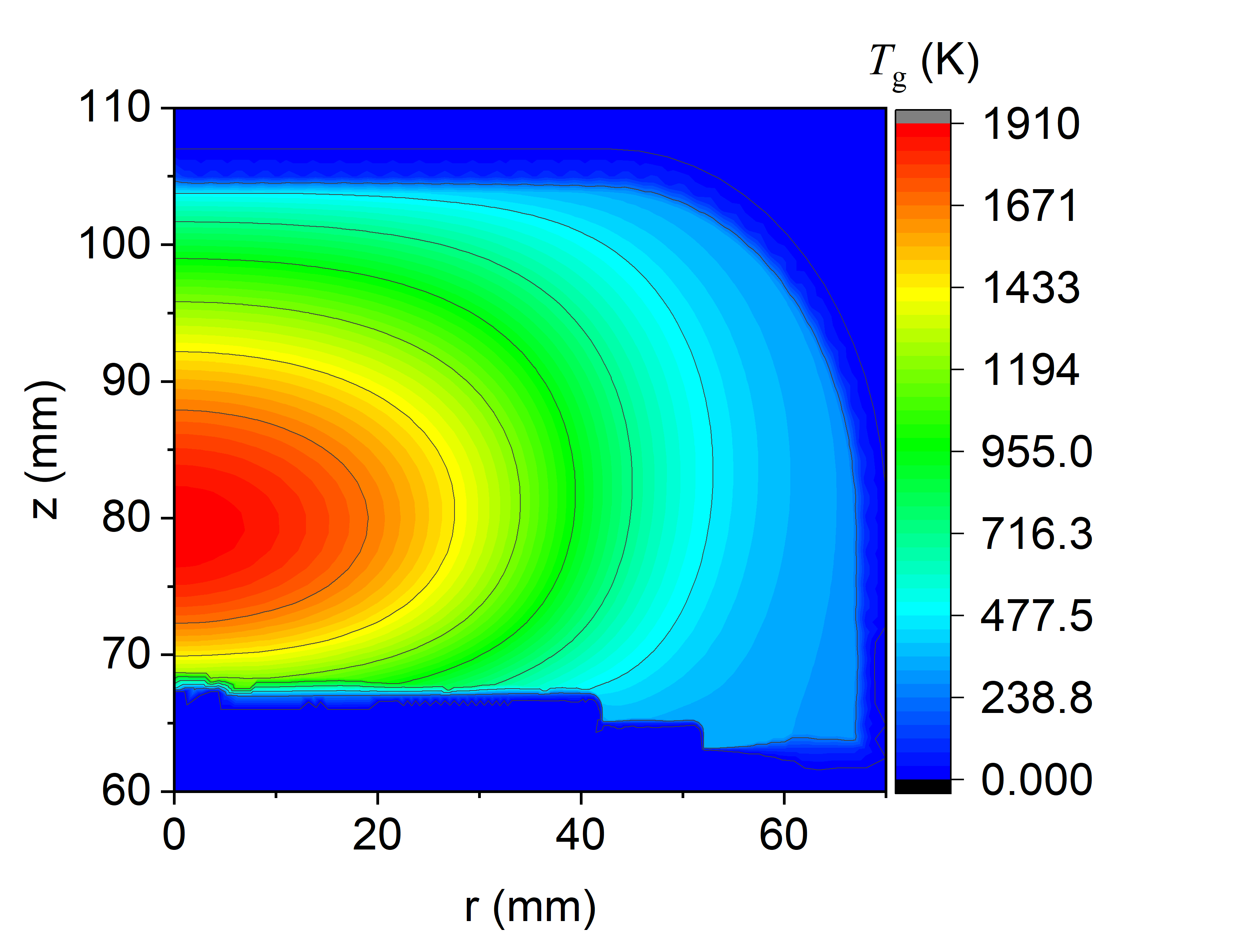}
        \caption{Gas temperature distribution} 
    \end{subfigure}
    \begin{subfigure}{0.32\textwidth}
        \includegraphics[width=\linewidth]{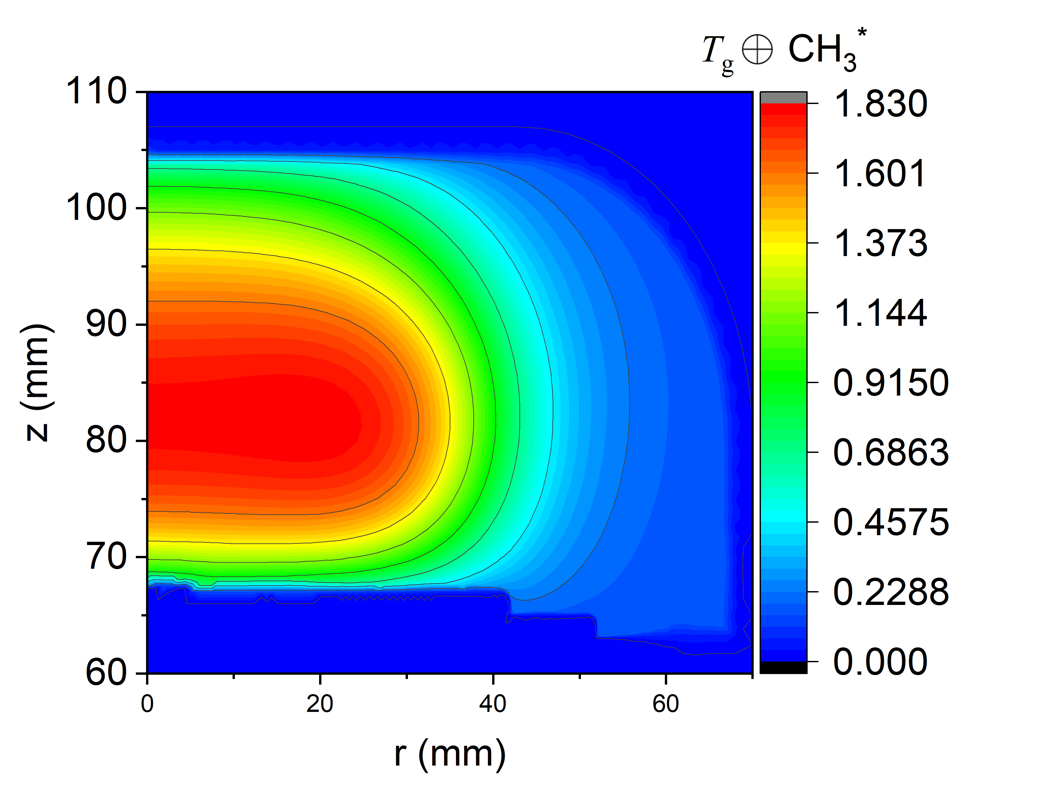}
        \caption{$T_g$ $\oplus$  CH$_\text{3}^{\bullet}$ distribution}
   \end{subfigure}
    \vspace{-0pt}
\caption{Contour plots representing the 2D spatial distribution of various plasma parameters across the reactor. }
\label{fig:40_Torr}
\end{figure*}

A simplified model of neutral species interaction was developed to describe the spatial evolution of C$_{x=1,2}$H$_{y=0-3}$  radicals within the chamber volume and their fluxes to the surfaces.\cite{Dandy_Coltrin1995, Gorbachev2022} The whole set of gas phase reactions considered is presented in Table  \ref{table:reactions}. Surface recombination of charged species was set to 1, and all carbon containing radicals are lost on the quartz dome walls with unity probability. 

Electric field distribution was calculated based on the Helmholtz equation given as:
\begin{equation}
    \nabla \times (\nabla \times \vec{E}) - k_0^2 \left( \varepsilon_r - \frac{j \bm{\cdot} \sigma_n}{\omega \bm{\cdot} \varepsilon_0} \right) \vec{E} = 0,
\end{equation}
where $k_0 = \omega \sqrt{\varepsilon_0 \mu_0}$ is the wave number in space. $\omega$ is the frequency, $\varepsilon_0$ is the absolute permittivity and $\mu_0$ is the absolute permeability of free space, $\varepsilon_r$ is the relative permittivity of the medium. $\sigma_n$ is the conductivity of the medium that gives rise to ohmic losses defined by the term $(j \bm{\cdot} \sigma_n)/(\omega \bm{\cdot} \varepsilon_0$). Electric conductivity and dielectric permittivity provide coupling between plasma and Helmholtz equation through the plasma frequency and electron transport frequency.

A heat transfer equation was used to couple the process of gas heating and temperature dependent chemical reaction with electromagnetic wave energy absorption in plasma. The heat balance equation can be written as:


\begin{equation}
    \rho_n C_p (\vec{u} \bm{\cdot} \nabla T) + \nabla \bm{\cdot} \vec{q} = Q_{mw}
\end{equation}

where $\rho_n$ is the fluid density $(kg/m^{-3})$, $C_p$ is the fluid heat capacity at a constant pressure $(J/kgK)$, $\vec{u}$ is the velocity vector responsible for convection in fluid $(m/s)$. In a continuous medium, conductive heat flux $\vec{q}$ is proportional to the temperature gradient, hence  $\vec{q} = -k\nabla T$, where $k$ is the thermal conductivity $(W/mK)$. $Q_{mw}$ is the heat source from microwave energy. 

Finally, gas inlet and outlet were considered in the model to take into account the process of CH$_\text{4}$ and H$_\text{2}$ neutrals renewal due to gas flow. Navie-Stocks equation given below was solved to obtain the spatial evolution of gas neutrals within the chamber volume.

\begin{equation}
    \rho_n (\vec{u} \bm{\cdot} \nabla) \vec{u} = \nabla \bm{\cdot} (-pI+\tau) + \vec{F}
\end{equation}

where $p$ is the pressure $(Pa)$, $I$ is the identity tensor, $\tau$ is the viscous stress tensor $(Pa)$, and $\vec{F}$ is the volume force vector $(N/m^3)$.

\subsubsection{$T_g$ and methyl radical distributions}


Fig.~\ref{fig:40_Torr} summarizes the computational results of the plasma model for the input reactor parameters detailed in Section~\ref{sec:Experiment}: (a) and (b) are contour plots representing the 2D spatial distribution of the methyl radical
CH$_\text{3}^{\bullet}$ and gas temperature $T_g$, respectively. 
Noting that the regions of maximum CH$_\text{3}^{\bullet}$ concentration and maximum $T_g$ do not overlap, Fig.~\ref{fig:40_Torr}(c) is the convolution of distributions (a) and (b): $T_g$ $\oplus$  CH$_\text{3}^{\bullet}$. This is done because, a priori, the location of the reaction zone (region of the plasma where diamond nucleation and subsequent grain formation occurs) is not known, estimates of the reaction zone are made based on:\\ 
($i$) the region of maximum CH$_\text{3}^{\bullet}$ concentration from Fig.~\ref{fig:40_Torr}(a);\\
($ii$) the region of maximum gas temperature from Fig.~\ref{fig:40_Torr}(b);\\
($iii$) the region where the concentration of CH$_\text{3}^{\bullet}$ and gas temperature are both optimal favoring the desired reactions: we estimate this region as the maximum of the convolution of CH$_\text{3}^{\bullet}$ concentration and $T_g$ from Fig.~\ref{fig:40_Torr}(c).

The respective gas temperatures obtained: $T_g^{CH_3^{\bullet}}$, $T_g^T$ and $T_g^{\oplus}$ corresponding to the maxima of CH$_\text{3}^{\bullet}$, $T_g$ and $T_g$ $\oplus$  CH$_\text{3}^{\bullet}$, respectively, are then used to re-calculate the activation energy from Eq.~\ref{eq:activation_energy}. 
The new values are plotted together with the activation energies calculated using $T_s$ 
in Fig.~\ref{fig:Ea}. 
Each of the four E$_\text{a}$ values, (obtained using $T_s$ and three estimates for $T_g$) is presented here with an error bar that originates from the linear fitting of the Arrhenius plot ln$(k)$ vs $1/T$ to obtain the term ln$(A)$ in Eq.~\ref{eq:activation_energy}.

\section{\label{sec:Discussion}Discussion}

\subsubsection{Comparison with SCD formation}

$E_a$ was reported to range from 13.6 to 22.5 kcal/mol for the formation of SCD\cite{Kondoh1993,Kang2000} and 0.98 kcal/mol for graphite.\cite{Fedoseev1979} 
 The $E_a$ values for nanodiamond films 
 calculated using $T_s$ range from 7.87 to 11.29 kcal/mol, as shown in Fig.~\ref{fig:Ea}. This range matches very well with the values 6--10 kcal/mol reported in the literature for nanodiamond formation,\cite{Brazil2014} where $E_a$ was always calculated using the substrate temperature. 
The revisited $E_a$ calculation gives a higher range of values: 13.5 $\pm$ 1.5 kcal/mol for $T_g^{CH_3^{\bullet}}$, 17.5 $\pm$ 2.5 kcal/mol for $T_g^T$ and 15 $\pm$ 2 kcal/mol for $T_g^{\oplus}$. These values are well within the characteristic range of SCD growth kinetics.

SCD ($sp^\text{3}$ carbon) formation is linked with the presence of CH$_\text{3}^{\bullet}$ in the plasma coupled with a high atomic H content to keep CH$_\text{3}^{\bullet}$ saturated with $\sigma$-bonded H terminations important for subsequent pure sp$^\text{3}$ phase formation and growth. Butler $et$ $al.$ \cite{Mankelevich2008_SCD} present a numerical simulation of the plasma used for SCD formation. Because of the high $T_g$ in the plasma core ($\sim$2900 K) achieved by high operating pressures of 150 Torr, molecular species are broken down to elementary parts and CH$_\text{3}^{\bullet}$ is distributed as a ring towards the outer cooler edge of the plasma. The optimal combination is when CH$_\text{3}^{\bullet}$ concentration is on the order of 10$^\text{13}$ cm$^\text{-3}$ at $T_g \sim 1200-1500$ K. This reaction zone is exactly where the SCD substrate is placed consuming the incoming flux of CH$_\text{3}^{\bullet}$. Following the hydrogen abstraction reactions, C atoms are added to the SCD substrate homoepitaxially, replicating the existing diamond structure. 

When compared to the described optical SCD growth conditions, our numerical simulation results illustrate that under 40 Torr operating condition the core of the plasma is much cooler with $T_g \sim 2000$ K. As in the SCD plasma case, the region for maximal CH$_\text{3}^{\bullet}$ concentration does not coincide with the maximal $T_g$ region. In fact, the maximal CH$_\text{3}^{\bullet}$ concentration is found to be at $T_g \sim 1400$ K, i.e. lying in the same temperature range as for the SCD plasma case. Because of the lower core temperature, the high CH$_\text{3}^{\bullet}$ concentration on the order of $~10^{14}$ cm$^{-3}$ is comparatively more evenly distributed throughout the plasma and the maxima of this distribution is not just concentrated on the edge. This makes sp$^\text{3}$ formation possible throughout the plasma 
giving rise to nucleation of diamond in the gas phase. 

Because of the low pressure though, the nucleated grain does not reach the critical size after which it can keep growing as a stable crystal. Thus, the process of renucleation begins giving rise to small crystal sized polycrystalline diamond. 
If now pressure is increased without changing any other parameters, the power density changes, the environment is no more in the metastable regime of diamond growth and the samples thus obtained are graphitized soot.\cite{Dynamic_Graphitization} Instead, if the pressure is increased and the power is reduced to balance the power density, it creates a suitable plasma that can sustain the growth of a diamond grain as it grows larger and hence the renucleation step is eliminated here.

\subsubsection{NCD film on SCD substrate}

The reasoning of renucleation due to low pressure was also supported by additional experimental observation. Synthesis parameters for NCD were applied to deposit a sample on an optical grade SCD substrate. Even though the C atoms from the plasma have a stable structure to replicate on and grow as a single crystal, the resulting final product was a highly renucleated NCD film. The resulting NCD on SCD film is shown in Fig. \ref{fig:NCD_on_SCD}. The SEM micrographs illustrate ballas-like structure with grain sizes comparable to the ones obtained on Si. This result is further confirmed by Raman spectroscopy before (SCD substrate) and after (NCD film) deposition. When the same reactor when operated at higher pressures (120 to 400 Torrs) and maintained at a constant substrate temperature of $\sim$1200 K (achieved by decreasing power and active substrate cooling), where the SCD substrate is directly subject to the ring shaped flux of high CH$_\text{3}^{\bullet}$ concentration, resulted in high quality single crystal homoepitaxy.\cite{Muehle2017} These results highlight the crucial importance of plasma kinetics on the final diamond product (NCD vs. SCD).

\begin{figure}
    
    \begin{subfigure}{0.23\textwidth}
       \includegraphics[width=\linewidth]{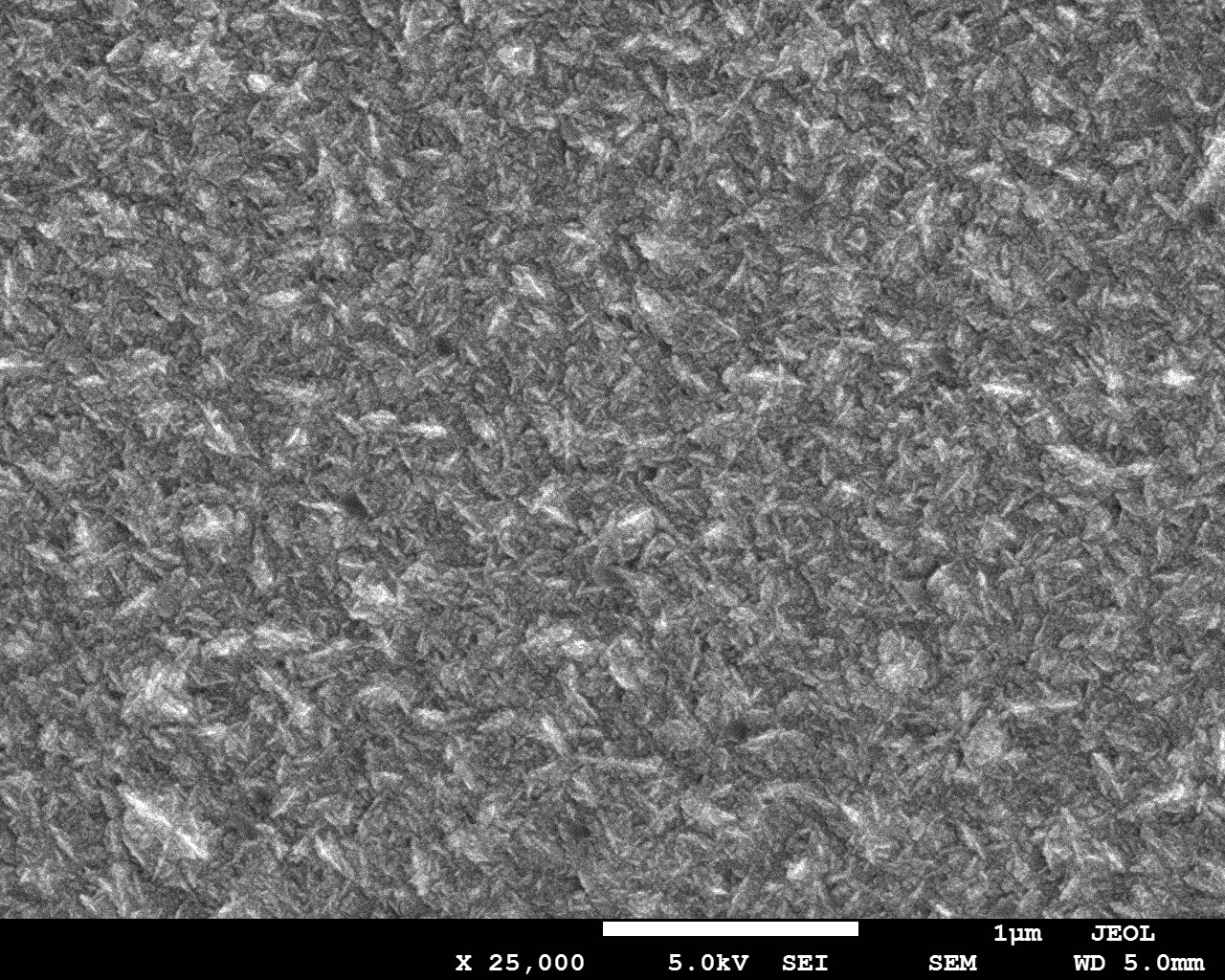}     
     \label{fig:NCD_on_SCD_SEM1}   
    \end{subfigure}
    \begin{subfigure}{0.23\textwidth}
       \includegraphics[width=\linewidth]{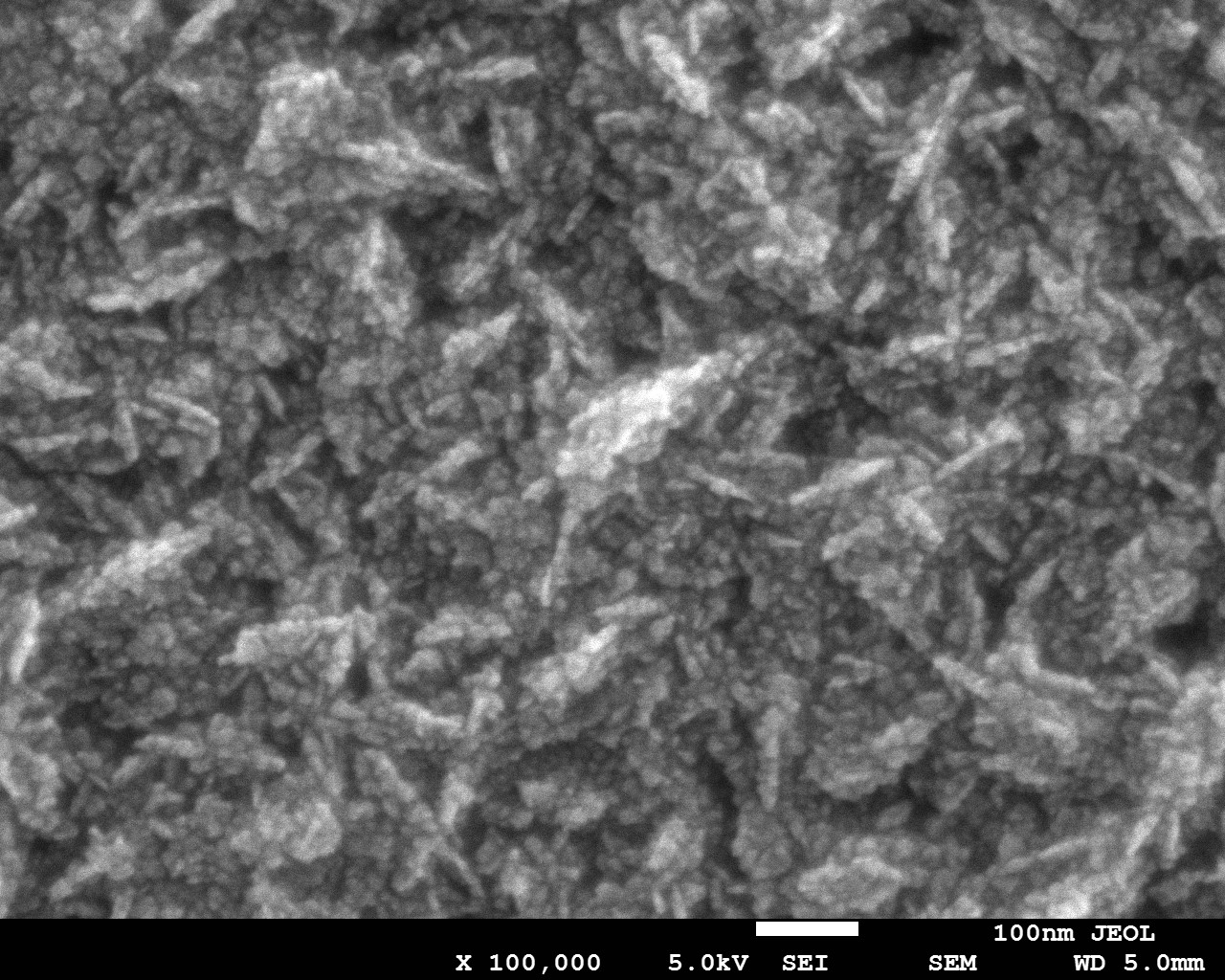}     
     \label{fig:NCD_on_SCD_SEM2}   
    \end{subfigure}

     \begin{subfigure}{0.45\textwidth}
        \includegraphics[width=\linewidth]{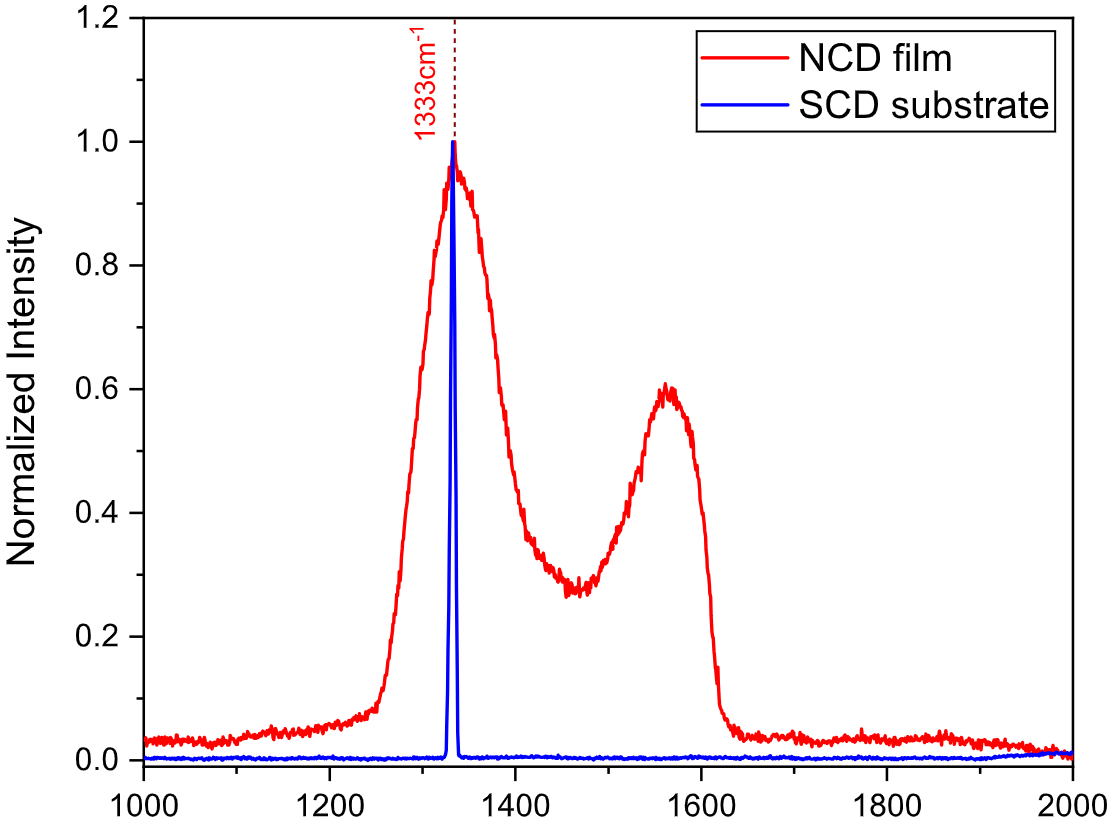}      
     \label{fig:NCD_on_SCD_Raman}
    \end{subfigure}
 
\caption{(top) SEM images, and (bottom) Raman spectrum of NCD film deposited on SCD substrate via in-plasma collection.}
\label{fig:NCD_on_SCD}
\end{figure}

\subsubsection{Proposed Mechanism}

\paragraph{Nanodiamond nucleation:}

In the gas phase, the particles precipitate from the supersaturated vapor into diamond crystals. In a H$_\text{2}$/CH$_\text{4}$ plasma, carbon can condense into both $sp^\text{2}$ and $sp^\text{3}$ phases from the gas phase. The preferential formation of diamond over graphite is usually explained by etching of $sp^\text{2}$ phase by atomic H but Hwang $et$ $al.$ \cite{Hwang1996} stated that H etching is not thermodynamically favored and instead proposed the capillary effect as the major effect behind $sp^\text{3}$ nucleation. The capillarity is expressed by Laplace-Young equation as

\begin{equation} P_2-P_1=2\sigma/r,
\end{equation}
where $P_2$ and $P_1$ are the pressures inside and outside the nucleus, respectively, of radius $r$ and surface energy $\sigma$. Given the mass density of diamond is larger than that of graphite, for the same number of C atoms precipitating into $sp^\text{2}$ and $sp^\text{3}$ phase forming a nucleus, the radius of this nucleus corresponding to $sp^\text{3}$ phase would be smaller than for $sp^\text{2}$ phase. Hence, for a given global gas pressure in the plasma of a low-pressure CVD reactor, $P_1$, the local pressure inside the nucleus, $P_2$ of $sp^\text{3}$ phase would be higher than that of $sp^\text{2}$ phase making it comparable to the pressure for making diamond more stable than graphite (local high pressure high temperature, HPHT condition). 

Based on this explanation, it can be inferred that as the diamond nucleates and grows to a certain size, the local pressure being inversely proportional to the radius is no longer high enough to stabilize this larger crystal. The crystal then stops growing and a new nucleus starts to form, hence the high renucleation rate in our low-pressure CVD environment. The gas phase growth of the diamond nanoparticle and the subsequent renucleation depends on the gas pressure. Lower the gas pressure, smaller the nanodiamond particles and higher the renucleation rate. For our conditions, the $r\sim10$ nm. As the operating pressure drops, nanodiamond stops nucleating and growing due to the decreasing gas temperature as $T_g$ is a strong function of the operating pressure. 

\paragraph{Nanodiamond particle and film growth:}

After subsequent renucleations, the condensed particles can no longer remain suspended in the plasma and drop down onto the substrate. This is illustrated by the schematic shown in Fig. \ref{fig:Nucleation_Schematic}.
In a recent work by Xiong $et$ $al.$,\cite{Xiong2022} it was noted that for nanometer-sized particles, gravity does not play a significant role. 
There is trapping in the plasma occurring out of the balance between the electrostatic force and opposing forces such as the drag of the flowing neutral gas, the ion drag, and the thermophoretic force. It was shown that sub-10 nm Si particles formed in a radiofrequency reactor due to nuclei being temporarily confined in a trap formed by electrostatic potential. Those particles only escaped after they grew to a size at which the increasing drag force was imparted by the flowing gas carrying the particles out. Because, generally, nanoparticles are expected to charge in plasmas a similar mechanism could be at play in our system. This would explain particles grow as they renucleate while being suspended in the gas phase until drag forces can take over to expel the cluster out of the plasma.

\begin{figure}[]
\centering
\includegraphics[width =0.5\textwidth]{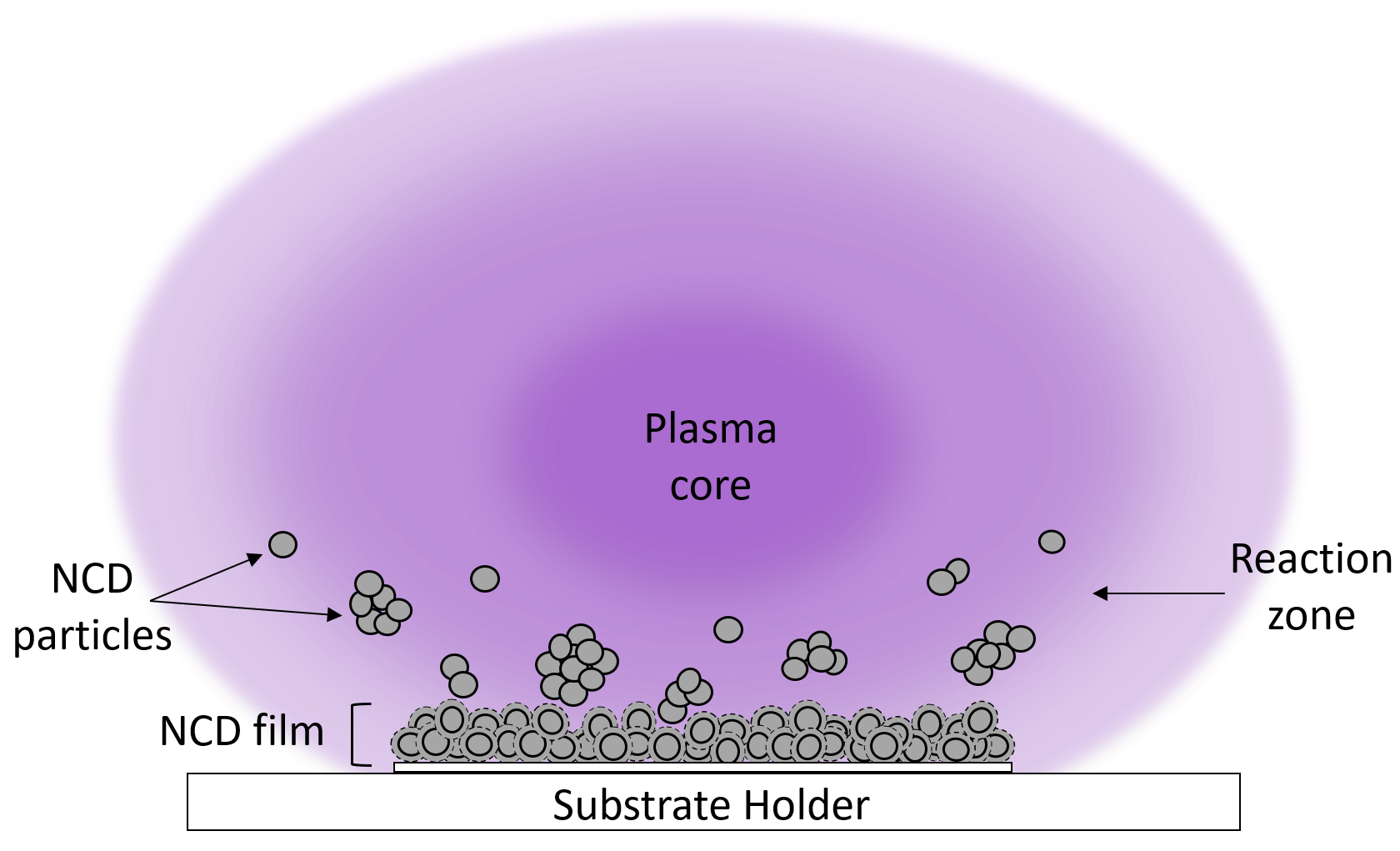}
\caption{Schematic of the proposed nucleation and growth of NCD film.} 
\label{fig:Nucleation_Schematic}
\end{figure}

This is further hypothesized that the main expelling force is of thermophoretic nature. There were two important references to note: Tsugawa $et$ $al.$\cite{Tsugawa2010} and Tallaire $et$ $al.$\cite{loose} In both cases, very similar sub-100 Torr plasma conditions were explored. In Tsugawa $et$ $al.$,\cite{Tsugawa2010} nanodiamond film was obtained at the substrate temperature less than $100\degree$C and in Tallaire $et$ $al.$,\cite{loose} loose nanodiamond film was obtained at the substrate temperature less than $500\degree$C. In either case, nanodiamond was drawn onto the substrate when the substrate was cooled significantly compared to the usual substrate growth temperatures.
Low $T_s$ would mitigate the substrate surface kinetics, indirectly highlighting the only other synthesis route through gas phase kinetics. In the present study, the substrate could not be cooled to $<500\degree$C and gas and substrate kinetics were instead divorced by physically separating the plasma and the collecting substrate through an orifice in the molybdenum plate (see Fig.~\ref{fig:Setup}).

Under conventional conditions where NCD or UNCD films, strongly adhering to the substrate, are grown on a seeded substrate placed into cooler plasma zone at temperatures 800-900 $\degree$C, 
both gas and surface kinetics play equally important role. Particles formed in the gas phase drop on the substrate, thereby enabling fast renucleation, while their regrowth on the substrate simultaneously takes place, thereby enhancing overall film growth.

\section{\label{sec:Conclusion}Conclusion}

Barbosa $et$ $al.$, \cite{Brazil2014} reviewed a consolidation of reports from which, based on the activation energy analysis, it was concluded that 
physico-chemical formation of SCD or MCD (faceted) is different from the formation of NCD and UNCD (ballaslike).
Our work presented new findings that point out that the formation of any kind of diamond (single crystal, faceted, ballaslike) is essentially the same and is governed by CH$_\text{3}^{\bullet}$ reaction as a limiting step. The main difference is that the formation of single crystal and faceted diamond is dominated by the substrate surface kinetics, while ballaslike nanodiamond formation takes place in the plasma , i.e. the nucleation, growth and renucleation is dominated by the gas phase kinetics. High renucleation happens because the conditions are not conducive to the formation of a single growing crystal. When the activation energy of SCD and MCD, calculated using the substrate temperature, was compared with that of NCD, calculated using the gas temperature, full agreement between the energies was found. Hence, high renucleation rate for NCD is not the rate-limiting step. 

The comparison of crystal size, morphology, and activation energy illustrated that the samples produced in the present study are directly comparable with the ballaslike nanodiamond materials widely reported in the literature. This warrants the extension of any results and hypotheses from the current work to nanodiamond materials reported elsewhere. This includes the call for reconsidering the importance of the gas phase NCD nucleation.



The exact reaction/formation belt in the plasma where nanodiamond particles (re)nucleate and grow remains unclear. From calculating E$_\text{a}$ using three gas temperatures estimates, the ideal conditions for nanodiamond formation and growth are where the zones of CH$_\text{3}^{\bullet}\sim 10^{13}-10^{14}$ cm$^{-3}$ and $T_g\sim 1200-1500$ K overlap. Finding exact interplays between various reactive species, gas temperature, nucleation and formation of sp$^2$ and sp$^3$ phases in the gas phase are the subjects of further studies.




\section*{Acknowledgement}
We would like to thank Dr. Mark J. Kushner (University of Michigan) and Dr. Rebecca Anthony (Michigan State University) for providing valuable insights into the subject. We are grateful to the U.S. Department of Energy, Office of Science, Office of Fusion Energy Sciences, Award No. DE-SC0023211,
and National Science Foundation, Division of Chemical, Bioengineering, Environmental and Transport Systems, Award No. 2333452 for the financial support.

\section*{Data Availability}
The data that support the findings of this study are available from the corresponding author
upon reasonable request.

\bibliography{Manuscript}
\end{document}